\documentclass[journal]{IEEEtran}
\IEEEoverridecommandlockouts
\usepackage{cite}
\usepackage{xcolor,soul,framed} 
\usepackage{amsmath,amssymb,amsfonts}
\usepackage{algorithmic}
\usepackage{graphicx}
\usepackage{textcomp}
\usepackage{siunitx}
\usepackage{booktabs}
\usepackage{float}
\usepackage{dblfloatfix}
\usepackage{array}
\usepackage{mdwmath}
\usepackage{mdwtab}
\usepackage{eqparbox}
\usepackage{url}

\usepackage{epsfig}
\usepackage{caption}
\usepackage{algorithm2e}
\usepackage{hyperref}
\hypersetup{
colorlinks=true,
linkcolor=black,
citecolor=black,
urlcolor=blue}
\author{Josef~Danial,~\IEEEmembership{Student~Member,~IEEE,}
        Debayan~Das,~\IEEEmembership{Student~Member,~IEEE,}
        Santosh~Ghosh,~\IEEEmembership{Member,~IEEE,}
        Arijit~Raychowdhury,~\IEEEmembership{Senior~Member,~IEEE}
        and~Shreyas~Sen,~\IEEEmembership{Senior~Member,~IEEE}}

\begin{document}
\bstctlcite{IEEEexample:BSTcontrol}
\title{\texttt{SCNIFFER}: Low-Cost, Automated, Efficient Electromagnetic Side-Channel Sniffing}
\maketitle
\begin{abstract}
Electromagnetic (EM) side-channel analysis (SCA) is a prominent tool to break mathematically-secure cryptographic engines, especially on resource-constrained IoT devices. Presently, to perform EM SCA on an embedded IoT device, the entire chip is manually scanned and the MTD (Minimum Traces to Disclosure) analysis is performed at each point on the chip to reveal the secret key of the encryption algorithm. However, an automated end-to-end framework for \textcolor{black}{EM leakage localization}, trace acquisition, and attack has been missing. This work proposes \texttt{SCNIFFER}: a low-cost, automated EM Side Channel leakage SNIFFing platform to perform efficient end-to-end Side-Channel attacks. Using a leakage measure such as TVLA\textcolor{black}{, or SNR}, we propose a greedy gradient-search heuristic that converges to one of the points of highest EM leakage on the chip (dimension: $N \times N$) within $O(N)$ iterations, and then perform Correlational EM Analysis (CEMA) at that point. This reduces the CEMA attack time by $\sim N$ 
times compared to an exhaustive MTD analysis, and $>20\times$ compared to choosing an attack location at random. We demonstrate \texttt{SCNIFFER} using a low-cost custom-built 3-D scanner with an H-field probe ($< \$\textcolor{black}{500}$) compared to $>\$50,000$ commercial EM scanners, and a variety of microcontrollers as the devices under attack. The \texttt{SCNIFFER} framework is evaluated for several cryptographic algorithms (AES-128, DES, RSA) running on both an 8-bit Atmega microcontroller and a 32-bit ARM microcontroller to find~\textcolor{black}{a point of high}  leakage and then perform a CEMA at that point. 
\end{abstract}

\begin{IEEEkeywords}
End-to-end EM SCA Attack, Low-Cost EM Scanning, Automated Framework, \texttt{SCNIFFER}
\end{IEEEkeywords}

\section{Introduction}
As the internet of things (IoT) continues to grow, security of many edge nodes has become critical. With many of these edge nodes being simple microcontrollers, side-channel attacks pose a powerful threat to their security. In the world of cryptography, side-channel attacks have long been identified as a threat to the security of computing and communication systems attempting to provide confidentiality and integrity of sensitive data, since the introduction of Differential Power Analysis in~\cite{kocher_differential_1999}. By analyzing physical side-channel information, such as power consumption, timing, or electromagnetic emissions, cryptographic algorithms that are mathematically secure can be broken efficiently. 
\begin{figure}[!t]
  \centering
  \includegraphics[width=0.48\textwidth]{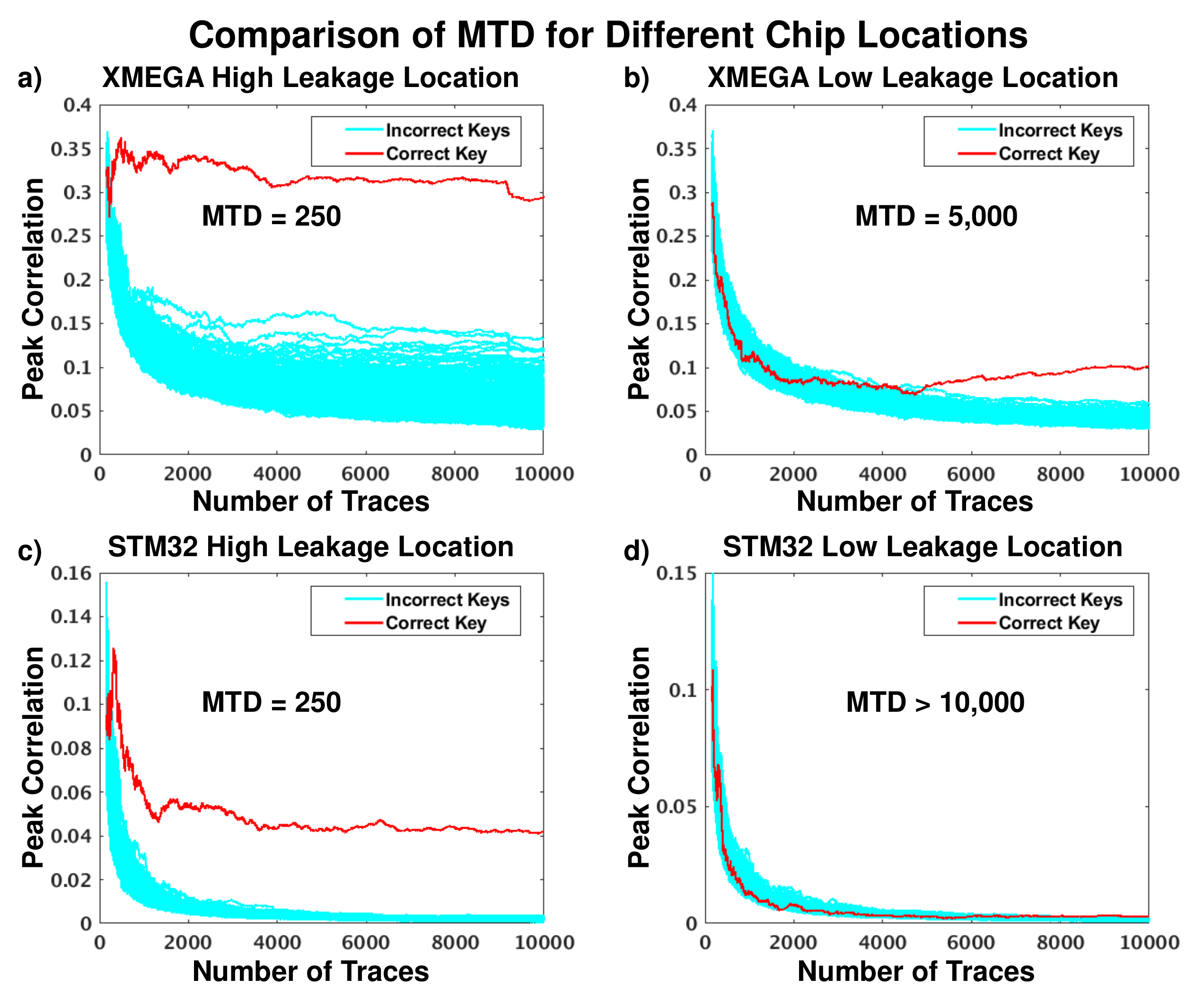}
  \caption{The difference in MTD between a CEMA attack at a point of high leakage vs. at a point of low leakage for both an 8-bit XMEGA microcontroller (a, b) and a 32-bit STM32F3 microcontroller (c, d). At a location of high leakage, the correct key separates in 250 traces for both microcontrollers, while a low leakage location requires $>20\times$ more traces on the XMEGA. At a low leakage location on the STM32F3, the key does not separate at all within 10,000 traces.}
  \label{EM_SCA_MTD_VAR}
\end{figure}

EM side-channel analysis (SCA) is a method of using the information found in the electromagnetic emissions of a cryptographic system to extract the secret key, compromising the security of such a system. Such attacks have been shown to be capable of actually extracting secret key information, as in~\cite{gandolfi_electromagnetic_2001} and~\cite{quisquater_electromagnetic_2001}. \textcolor{black}{These EM emissions originate from current consumption of an IC running cryptographic algorithms, which while flowing through the metal layers of an IC cause EM radiation as described in~\cite{das_stellar:_2018}. The EM emissions can either be caused by key-dependent operations or other operations, the key-dependent operations create key-dependent EM emissions, which contribute to the side-channel signal, while other operations contribute to algorithmic noise.} EM SCA attacks have successfully been used in the real world on PCs, shown in~\cite{genkin_stealing_2015} and~\cite{genkin_ecdh_2016}, and also on Smart Cards, in~\cite{matthews_low_2006}~\cite{kasper_em_2009}.
\textcolor{black}{One powerful and commonly used side-channel analysis technique is correlational electromagnetic analysis (CEMA).} In CEMA, EM measurements are taken while a cryptographic algorithm is executing on the target system (each measurement is known as a trace), and these traces are correlated with a leakage model, such as the Hamming Weight or Hamming Distance of data at a particular point in an algorithm~\cite{kocher_differential_1999}, under a hypothesis of a subset of the secret key. In a successful attack, the hypothesis that results in maximum correlation corresponds to the secret key. By attacking the hidden key incrementally, for example one byte at a time for AES-128, the entire secret key can be recovered, in orders of magnitude less time than brute force or other cryptanalysis methods.


\subsection{Motivation}
EM side-channel attacks, while powerful in that they are non-invasive and do not require any physical changes to the system being attacked, and benefit from allowing an attacker to choose the \textcolor{black}{location} with maximum information leakage (\textcolor{black}{SNR}), introduce a number of additional challenges compared to the power SCA attacks. Firstly, as the EM signals go through a power to EM transformation that reduces amplitude compared to the measurement noise floor, meaning more traces, \textcolor{black}{or more expensive measurement equipment may be needed}  to perform an attack. Secondly, unlike power attacks, EM attacks require attackers to choose the location of the attack in the system to capture the EM traces. This choice can have a drastic impact on the effectiveness and efficiency of an attack. As seen in Figure \ref{EM_SCA_MTD_VAR}, depending on where the EM probe is placed on a chip, the MTD for a CEMA attack can vary by $>20\times$, even for the small 9mm x 9mm Atmega and STM microcontrollers used as the target devices for this work. Current methods for determining the best location to perform CEMA are based on exhaustive search, simply performing a CEMA attack at most locations. Alternatively, it is also possible to choose an arbitrary location, and use as many traces as necessary to perform the CEMA. Practically, if the size of the system is \textcolor{black}{larger}, finding the correct location of the EM leakage becomes extremely challenging and requires scanning the entire chip/system. 

Given the limitations of present attack systems, in this work, we propose a low-cost, fully automated, end-to-end platform for performing efficient EM side-channel attacks. The core of this framework is a $\sim\$200$ 3-D printer, which we have \textcolor{black}{modified} to utilize as a low-cost EM scanner. \texttt{SCNIFFER} also uses a greedy gradient-search heuristic using a leakage measure,\textcolor{black}{ such as  test vector leakage assessment (TVLA), or SNR to quickly and automatically locate a point of high leakage}. Finally, once the point is determined, the proposed \texttt{SCNIFFER} framework performs the correlational or differential EM \textcolor{black}{analysis} (CEMA/DEMA) at this point. \textcolor{black}{While both CEMA and DEMA are possible attacks, throughout this work, we will demonstrate results with CEMA.} Such an automated low-cost attack platform significantly increases the threat surface for IoT devices.

\subsection{Contribution}
Specific contributions of this article are:
\begin{itemize}
\item Firstly, a fully-automated system for efficiently scanning a cryptographic chip and finding \textcolor{black}{a location of high leakage} to mount an end-to-end EM SCA attack is proposed. The entire attack set-up is extremely low-cost, owing to the custom-built EM scanner (\textcolor{black}{adapting} a $\sim\$200$ 3-D printer) used for mounting the attack, compared to the commercially available EM probe stations, which are very costly ($>\$50,000$). The system achieves 100$\mu$m spatial resolution, \textcolor{black}{and} has a scan range of 220mm $\times$ 220mm, and is easily replicable. (Section 3)

\item Secondly, a greedy gradient-descent heuristic is proposed which converges to \textcolor{black}{a point of high leakage} on an $N \times N$ chip within $O(N)$ iterations. This algorithm is evaluated with \textcolor{black}{both TVLA and SNR} as the measures of leakage. (Sections 4, 5)

\item Finally, the \texttt{SCNIFFER} attack is demonstrated on two different microcontroller architectures (8-bit XMEGA and 32-bit STM32F3), improving the number of traces required by $\sim 100\times$ compared to the traditional exhaustive search based attack. (Sections 5, 6)
\end{itemize}

\subsection{Paper Organization}
The remainder of the paper is organized as follows. Section 2 provides the background and summarizes the existing works on EM Scanning and side-channel attacks. In Section 3, the \texttt{SCNIFFER} framework is introduced and the low cost, custom-built EM scanning platform is presented. Section 4 describes two options for measuring leakage, signal amplitude and TVLA, and provides motivation for finding the point of highest leakage. In Section 5, the gradient-descent algorithm for efficiently determining \textcolor{black}{a point of high information leakage} is proposed. Next, Section 6 provides results of running the system on microcontrollers of varying architectures, cryptographic algorithms executed, and measures of leakage. Finally, Section 7 concludes the paper.

\section{Background and Related Work}
IoT devices have been successfully attacked using side channel attacks, for example CPA was used to extract encryption keys from Philips Hue smart lamps in~\cite{ronen_iot_2017}. EM side-channel attacks \textcolor{black}{were first proposed} in~\cite{agrawal_em_2003}, and share many properties with power side-channel attacks, however, can be performed at a distance, even up to one meter, as in~\cite{ramsay_tempest_2018}. One of the most powerful EM SCA attacks is CEMA, which is the straightforward application of Correlation Power analysis (CPA)~\cite{brier_correlation_2004} on EM traces.

However, to make these profiled and non-profiled EM SCA attacks more practical and real-time on any embedded platform/device, the trace capture and the attack needs to be automated and more efficient.

\texttt{SCNIFFER} can use several methods of assessing leakage, \textcolor{black}{for instance, simple signal magnitude, Test Vector Leakage Assessment (TVLA)~\cite{becker_test_2013}, or SNR~\cite{mangard_hardware_2004}}. In \textbf{TVLA}, two sets of traces are collected. In one set, both the key and plaintext used as input to the algorithm under test are kept fixed, and in the other the plaintext is varied randomly, while the key remains fixed. To assess the leakage, one then performs Welch's t-test for each time point of the trace. \textcolor{black}{Welch's t-test is given by $t = \frac{\Bar{X_1} - \Bar{X_2}}{\sqrt{\frac{s_1^2}{N_1} + \frac{s_2^2}{N_2}}}$, where $\Bar{X_1}, \Bar{X_2}$ are the sample means of the two sets, $s_1, s_2$ are sample standard deviations for the sets, and $N_1, N_2$ are the sizes of the sets. If the maximum t-value at a point is above 4.5, one can conclude leakage is present with 99.999\% confidence. Meanwhile, we consider the signal to noise ratio as defined in~\cite{mangard_hardware_2004}, to be $\textbf{SNR}=\frac{VAR[Q]}{VAR[N]}$, where $Q$ is the side channel leakage, and $N$ is the noise. Unlike TVLA, which does not guarantee exploitable leakage, SNR defined in this way can be directly related to the success rate of a CEMA attack~\cite{mangard_hardware_2004}.}

\textcolor{black}{Once \texttt{SCNIFFER} has chosen a point to attack, CEMA is used to recover the secret key. CEMA revolves around making hypotheses on secret values, then predicting the EM leakage of an intermediate variable based on the key. Measurements (traces) are taken while the device performs encryption, then the measurements are correlated with the predicted leakage for all hypotheses. The hypothesis that results in the largest correlation is taken as the guess for the secret value. The number of traces needed to recover the key in this way is then the minimum traces to disclosure (MTD). In this work, the secret values are the bytes of the AES key, and the intermediate variable is the first round sbox output, and Hamming Weight, that is, the number of 1's in the binary representation of this variable, is used as the leakage model of data at this point.}

Addressing the issue of finding where a chip leaks the most EM radiation has been investigated in\textcolor{black}{~\cite{liu_fully_2014}}, and~\cite{lomne_modeling_2011}. EM scanning with a focus on side-channel attacks, that is, determining where the most cryptographic information leaks within a chip has been addressed in~\cite{sauvage_electromagnetic_2008} and~\cite{heyszl_localized_2012}. However, such methods focus on observing the leakage over the entire chip, not efficiently finding the point or region of the maximum leakage. This causes these methods to take a long time and a majority of the time is spent collecting data that is unnecessary for an attacker. By creating a framework that minimizes this unnecessary data collection, EM side-channel attacks can be made more efficient, powerful, and practical, requiring far fewer traces to reveal the secret key of the cryptographic algorithm. Additionally, these platforms can be orders of magnitude more costly than the system proposed in this work, for instance the Riscure EM Probe station\textcolor{black}{~\cite{noauthor_em_nodate}} itself can cost $\sim\$50,000$, while the entire \texttt{SCNIFFER} system costs $<\$500$. \texttt{SCNIFFER} is the first fully-automated, efficient EM SCA attack framework and the system is described in the following section.

\begin{figure}[!t]
  \centering
  \includegraphics[width=0.48\textwidth]{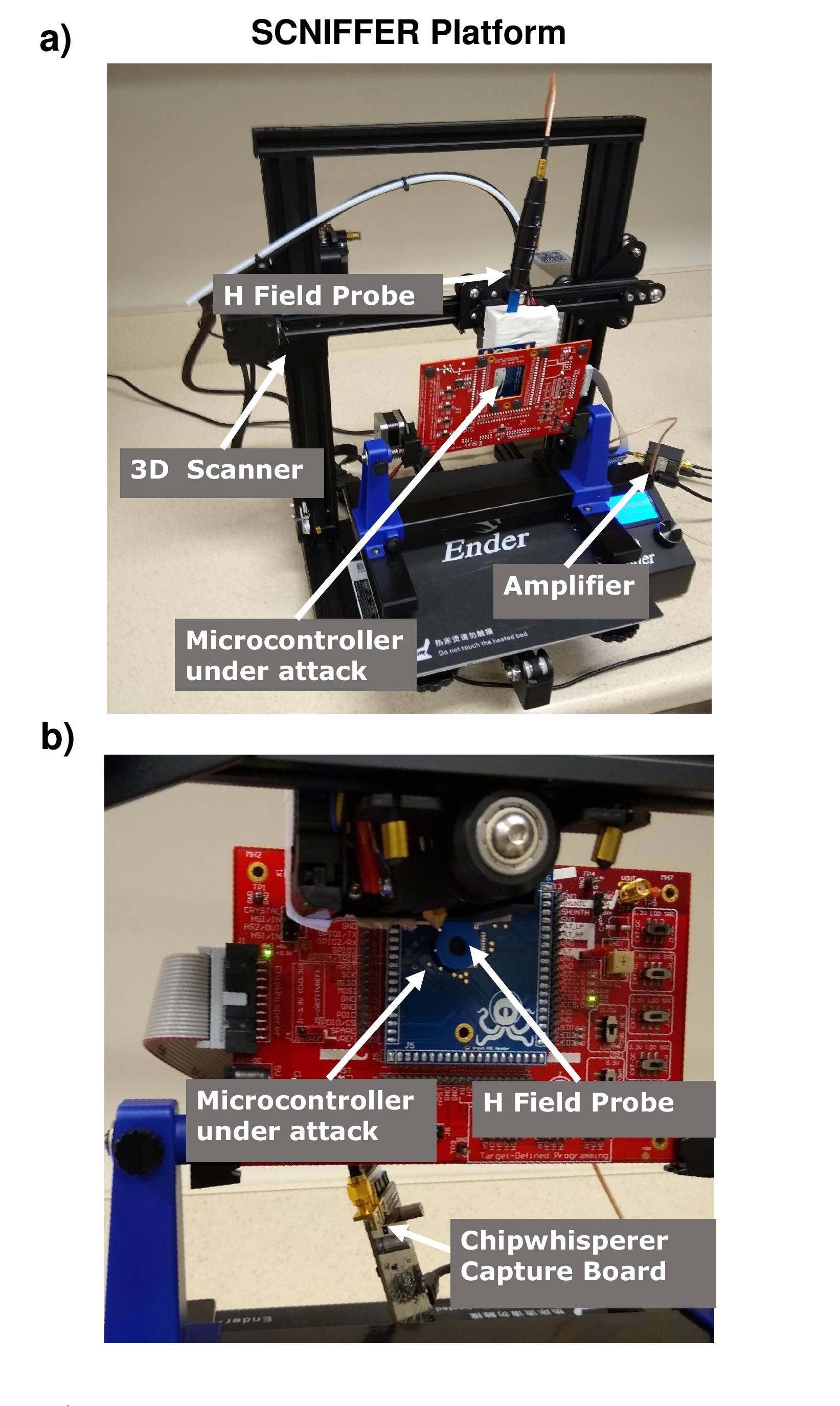}
  \caption{(a) The complete EM Scanning and trace capture set-up system, including the 3-D printer, Chipwhisperer system, EM probe, amplifier, and victim. (b) Close-up of scanner, showing probe and victim board.}
  \label{SCNIFFER_BLOCK_Diagram}
\end{figure}

\begin{table}[!t]
  \centering
  \includegraphics[width=0.48\textwidth]{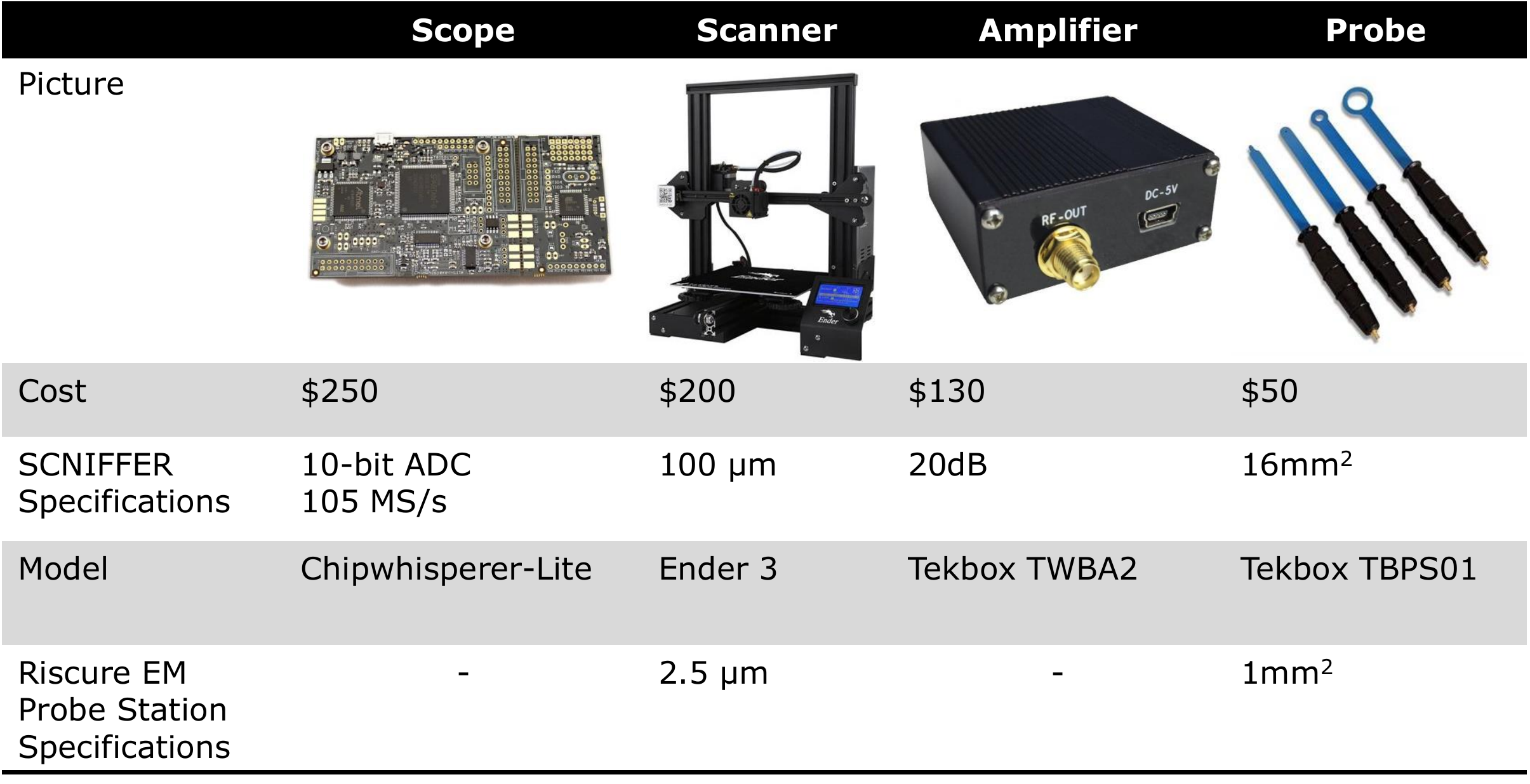}
  \caption{Summary of the main components of the \texttt{SCNIFFER} system, their costs, performance, and a comparison to Riscure's EM Probe Station.}
  \label{SCNIFFER_Table}
\end{table}

\begin{figure}[!t]
  \centering
  \includegraphics[width=0.48\textwidth]{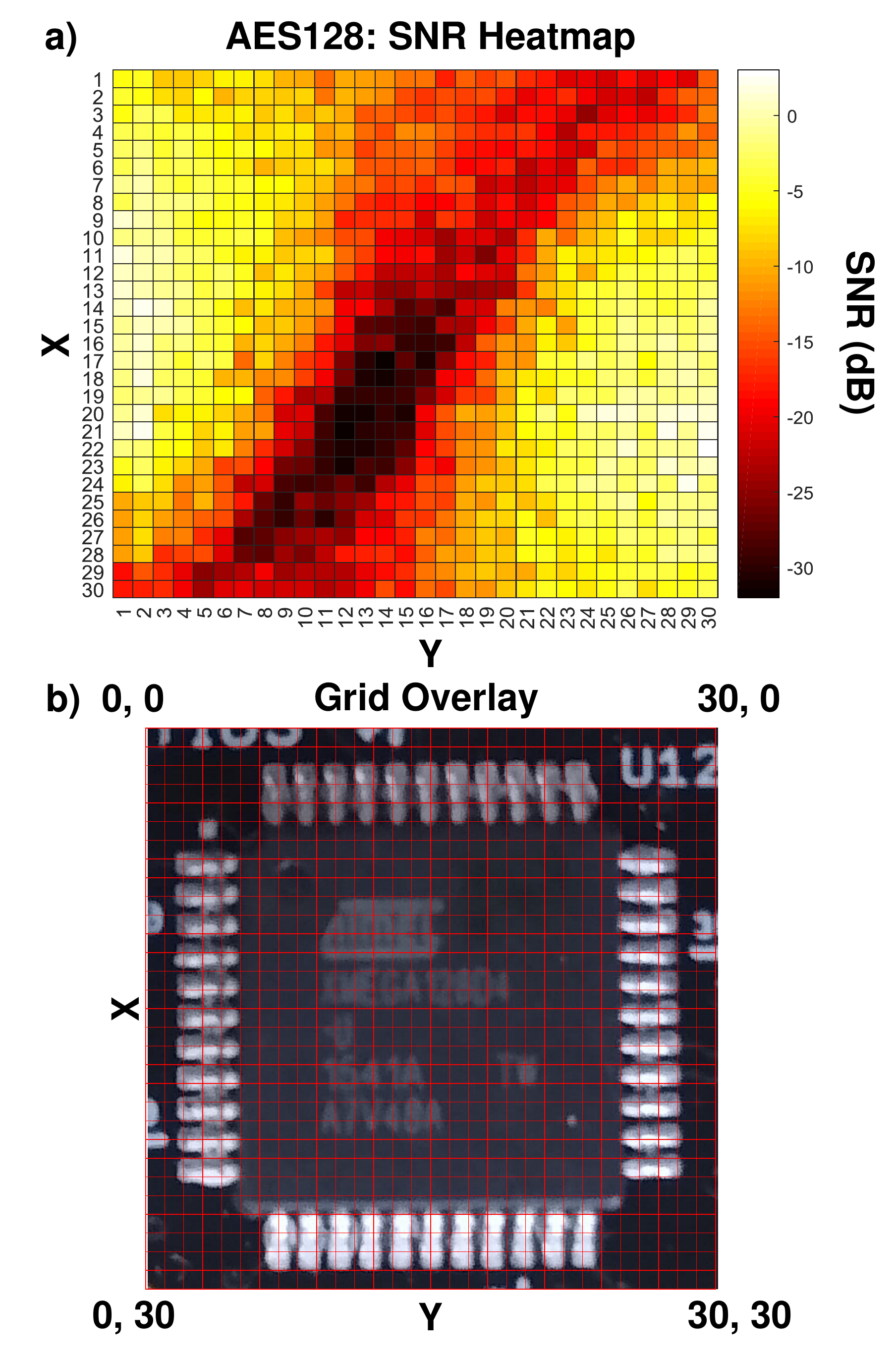}
  \caption{\textcolor{black}{(a) Heatmap of the SNR values obtained by performing a full $30 \times 30$ scan of the 8-bit target microcontroller.} (b) This shows the grid divisions where leakage measurements were performed. \textcolor{black}{1000 traces were used to compute the SNR values} at each point. The part of the target microcontroller board which leak the most information can be observed.}
  \label{SNR_Heatmap}
\end{figure}

\begin{figure}[!t]
  \centering
  \includegraphics[width=0.48\textwidth]{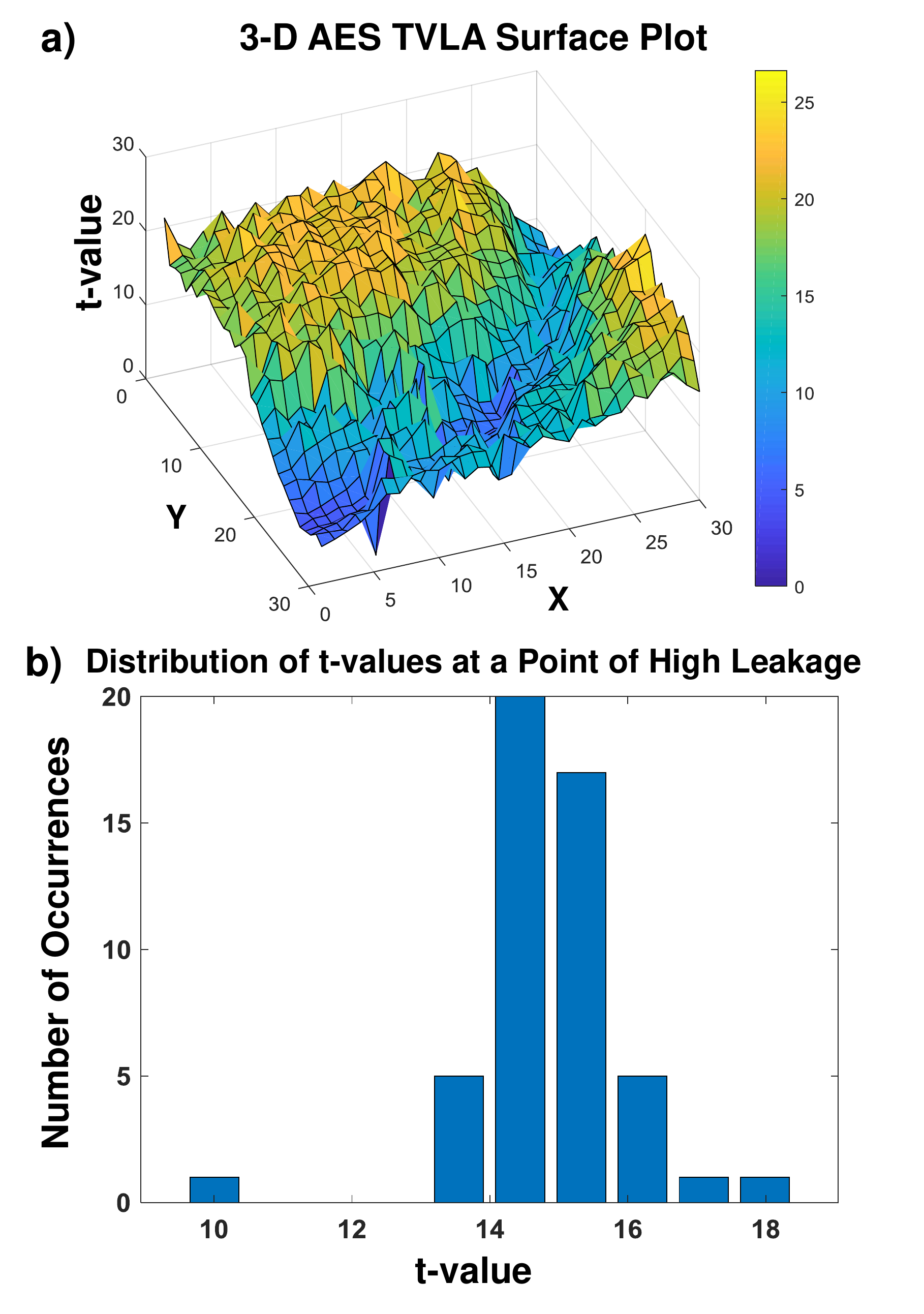}
  \caption{(a) TVLA Surface plot. Again, the surface is not smooth or monotonic, as there are many local minima and maxima, as in Figure \ref{SNR_Surface}(a). (b) Histogram of TVLA measurements at a single point. 50 TVLA measurements were made at a point of high leakage, each done as in (a), using 400 traces each. Given the distribution much wider seen in (b), the increased roughness of the surface in (a) can be explained. }
  \label{TVLA_Surface}
\end{figure}
\begin{figure}[!t]
  \centering
  \includegraphics[width=0.48\textwidth]{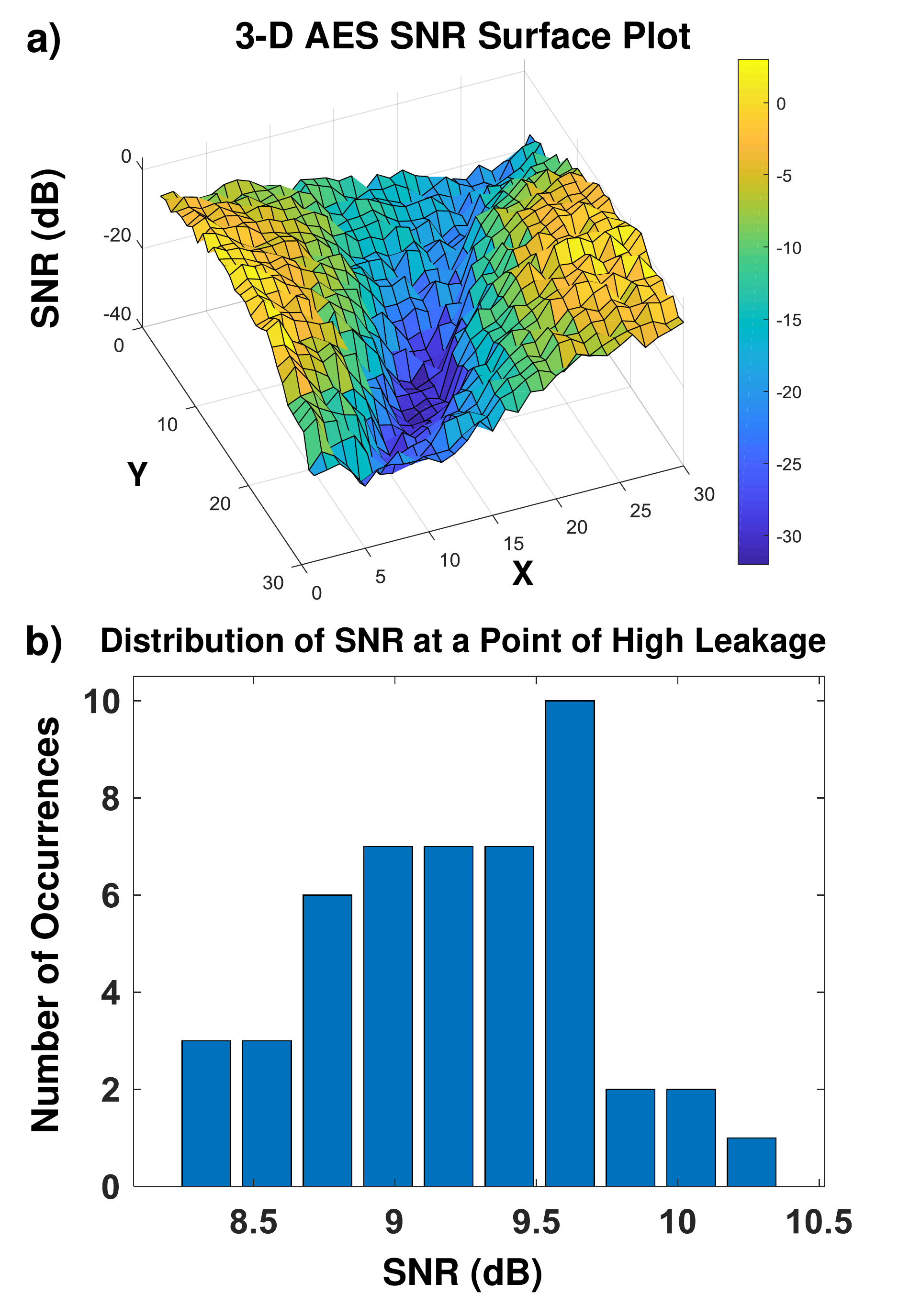}
  \caption{(a) \textcolor{black}{SNR surface plot of the same scan as Figure \ref{SNR_Heatmap}(a). Here it can be clearly seen that the surface is not smooth or monotonic, as there are many local minima and maxima. (b) Histogram of SNR measurements at a single point. 50 SNR measurements were made at 1 point. This distribution can explain some of the roughness of the surface seen in (a).} }
  \label{SNR_Surface}
\end{figure}

\section{\texttt{SCNIFFER}: Low Cost Automated EM Scanning}
The \texttt{SCNIFFER} system is designed for low cost and automation. In this section, we first describe the physical components that make up \texttt{SCNIFFER}, then discuss the automation aspect of the system.
\subsection{Low Cost EM Scanning Setup}
The scanning hardware consists of an Ender-3 3-D printer~\cite{noauthor_creality3d_nodate-1} with a \textcolor{black}{10mm loop diameter} H-field probe attached to the extruder, the Chipwhisperer~\cite{oflynn_chipwhisperer:_2014} platform for interfacing with the victim \textcolor{black}{(The CW309T-XMEGA mounted on the 308 UFO Target board)} and trace collection, an amplifier to amplify the EM probe output, and finally a PC to control both the 3-D printer and \textcolor{black}{the Chipwhisperer Lite  capture board}. While such EM scanning systems do exist, for instance, Riscure's EM Scanning Station, we chose to create such a system from scratch for the following reasons: 1) Commercial scanning systems (like Riscure~\cite{noauthor_em_nodate}) scanning station is orders of magnitude more expensive and 2) It is very straightforward to interface with the custom system to develop the scanning algorithm. \textcolor{black}{As seen in Table~\ref{SCNIFFER_Table}, the cost of a commercial scanner is orders of magnitude higher than \texttt{SCNIFFER}, and while it is hard to know if this price has been inflated by the selling company, it is reasonable for prices to be higher, as there are not many EM scanners on the market.}

To manipulate the probe, an Ender-3 3-D printer, running stock firmware was used. This model of printer has a minimum step size of 0.1mm, and can be controlled via a USB serial connection. It has a maximum movement speed of 180 mm/s, with a print area of $220mm\times220mm\times250mm$. The precision and speed offered by this 3-D printer are sufficient to complete a $50\times50$ scan of the $9mm\times9mm$ IC used in testing in an acceptable time. \textcolor{black}{Additional justification for the choice of printer, beyond the cost includes the ease of interfacing, the form factor, maintainability, and software support. The open source firmware used by this printer is well documented, and can be controlled through an exposed serial port, making interfacing very easy. The printer also has an open form factor that allows the probe and victim board to be mounted easily. While the durability and hardware support would not be as good as a commercial EM scanner, the simple construction and use of off-the-shelf components make maintenance straightforward. The software support is quite strong, being open source, and the printer is plug-and-play compatible with any device with a serial port.} The system is capable of performing a $30\times30$ scan of the chip in $\sim15$ minutes, and perform an amplitude scan in $\sim75$ minutes. The probe used is a commercial H-field probe for performing EMC measurements, and the signal is amplified before being passed to the Chipwhisperer capture board. \textit{While the probe used does not have extremely high spatial resolution, the probe resolution matches the scan resolution}, allowing heatmaps such as the one in Figure \ref{SNR_Heatmap}(a) to be created, and Chipwhisperer is able to capture enough information leakage for the target devices considered, leading to low MTDs when probed at appropriate locations, as seen in figure \ref{EM_SCA_MTD_VAR}, while still being low cost. \textcolor{black}{Even though this probe is on the larger side, the \texttt{SCNIFFER} platform is compatible with more sensitive probes and is expected to become more precise with such probes.} The complete system is shown in Figure \ref{SCNIFFER_BLOCK_Diagram}(a) showing the 3-D printer, the probe, Chipwhisperer system, and PC. The probe and victim IC are shown in detail in Figure \ref{SCNIFFER_BLOCK_Diagram}(b). The probe position can be controlled manually, through the 3-D printer controls, or programmatically through the serial connection to a PC, as it is in the \texttt{SCNIFFER} system.

The major cost savings in the \texttt{SCNIFFER} system come from using a low cost 3-D printer to control the probe, instead of a high cost motorized table. The total cost of the 3-D printer, probe and amplifier used in \texttt{SCNIFFER} is $\sim\$500$, which is a few orders of magnitude less expensive than many motorized tables by themselves, and nearly two orders of magnitude less expensive than systems such as Riscure's EM probe station $(\sim \$50,000)$. While more expensive scanners, probes and measurement systems could improve spatial and frequency resolution, such a system would only be available to very sophisticated attackers. As \texttt{SCNIFFER} aims to demonstrate practical, low-cost attacks are possible using systems two orders of magnitude cheaper than existing scanners, high-cost, high resolution components are not used. Table \ref{SCNIFFER_Table} summarizes these components, including their costs and performance \textcolor{black}{compared} to the Riscure system. 

\subsection{Automated EM Scanning}
Now that the \texttt{SCNIFFER} system's low cost hardware has been described, we move to the automated scanning and attack procedure. The basic premise of the automated system is to locate a point on the target device \textcolor{black}{where the chosen leakage measure is high} by using the scanning algorithm specified in Section 5, then automatically perform CEMA at this point. This removes the need for an expert to manually analyze example traces to choose a location for an attack.

During an attack, the probe is positioned at a location dictated by the intelligent scanning algorithm, then, the appropriate ADC phase for trace collection is determined by capturing traces at varying ADC phases, and the phase giving the largest average amplitude is chosen for further measurements at that particular point. \textcolor{black}{The signal is sampled at 29.48MHz, 4$\times$ the clock frequency of 7.37MHz, so clock edges are aligned to samples. The signal is amplified by the external amplifier, as well as the Chipwhisperer internal amplifier (set to a gain of 34.5dB), but no other prepossessing is performed.} Chipwhisperer is then used to capture traces for leakage measurement (through \textcolor{black}{SNR, TVLA} or other measures) and finally CEMA is performed at the \textcolor{black}{location found by the algorithm to have the highest leakage}. Example leakage measures tested with \texttt{SCNIFFER}, and the development of the intelligent scanning algorithm, along with detailed results are described in the following sections. 
\begin{figure}[!t]
  \centering
  \includegraphics[width=0.48\textwidth]{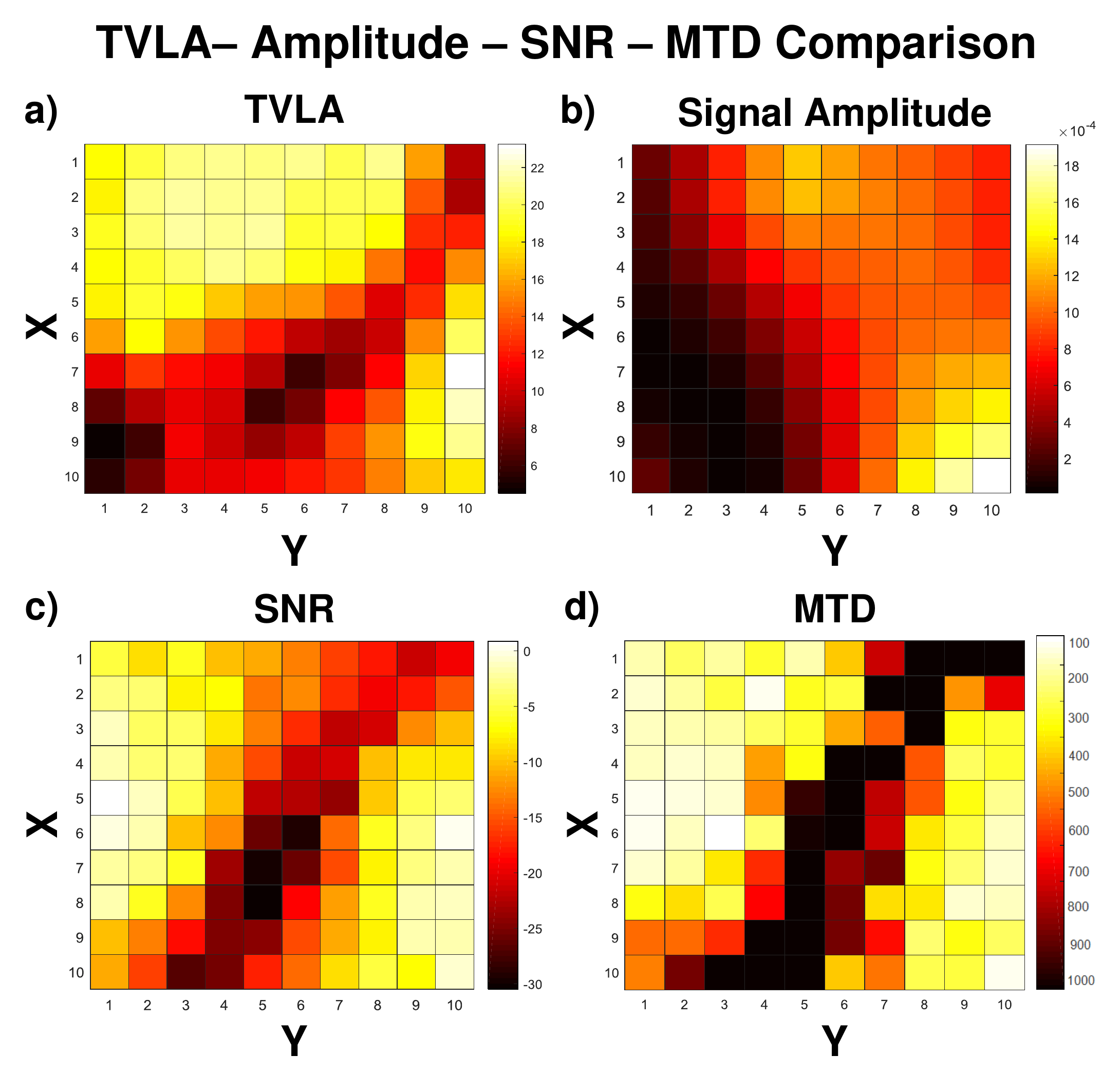}
  \caption{\textcolor{black}{10 $\times$ 10 heatmap of (a) TVLA values (b) signal amplitudes (c) SNR values and (d) MTDs. From these plots TVLA and SNR appear to correlate to MTD much better than the signal amplitude. While amplitude is easy to measure, it is clear that high amplitude of leakage does not necessarily correspond to high information leakage.}}
  \label{MTD_Compare}
\end{figure}
\section{Signal Leakage Measurement using \texttt{SCNIFFER}}
As the choice of probe location is a major factor in determining the number of traces needed to recover a key in CEMA as shown in Figure \ref{EM_SCA_MTD_VAR}, this location must be chosen intelligently. Currently, this is done by either exhaustive search of the entire chip, or by an expert evaluating sample EM traces at several locations, and choosing a location based on visual inspection \textcolor{black} {of the traces}. While an exhaustive search will certainly produce the best location to attack, it requires a large amount of time, especially for systems with a large initial MTD. Choosing a location based on visual inspection \textcolor{black} {of traces} may result in a location that can be attacked, however not necessarily the best in terms of MTD. Additionally, this method requires an expert to perform the inspection of traces. In this work, we aim to fully automate the process of choosing a location as an expert might, by looking at measures of leakage, and finding \textcolor{black}{a location with high leakage}. As with a manual choice, this location may not be the location corresponding to the lowest MTD, but should leak enough information to be attacked in a reasonable amount of time, \textcolor{black}{without the need for an expert}.

\texttt{SCNIFFER} is designed such that any measure of leakage can be used. For example signal amplitude, Test Vector Leakage Assessment (TVLA)~\cite{becker_test_2013}, \textcolor{black}{or SNR}  could be used, and the \texttt{SCNIFFER} platform will be able to converge \textcolor{black}{to a location where the leakage measure is high} in $O(N)$ measurements. We provide results using both \textcolor{black}{TVLA and SNR}, both described, and then compared in the following subsections.

\subsection{Signal Amplitude for Leakage Measurement}
\textcolor{black}{As motivation for why side-channel leakage measures must be used with \texttt{SCNIFFER} to locate low MTD locations, we measure the signal amplitude at each point of the victim chip, producing the heatmap seen in Figure \ref{MTD_Compare}(b). The amplitude was measured as the mean square amplitude of each trace, averaged across 10 traces. As can clearly be seen in that figure, the amplitude does not correlate to the MTD at all, as expected.}

\textcolor{black}{Hence, further results are shown using one of the two leakage measures explained in the following sections, TVLA and SNR. While these are the measures chosen for demonstrating \texttt{SCNIFFER}, they are by no means the best nor the only measures that can be used, as \texttt{SCNIFFER} does not rely on specific leakage type, only requires that the leakage correlate with the MTD. Determining the best measures of leakage in terms of the attack success rate and minimum number of traces required is a future research direction.}

\subsection{TVLA for Leakage Measurement}
\textcolor{black}{While signal amplitude is quick to measure, it has no relationship to side channel leakage. As the goal of \texttt{SCNIFFER} is to locate a position with high side channel leakage, amplitude is therefore not a good measure. A measure that does consider side channel leakage, and may be a better fit for \texttt{SCNIFFER} is TVLA}. While high t-values from TVLA may not necessarily imply a low MTD, it allows locations \textcolor{black}{where leakage is detected with high confidence  to be focused on.}
The TVLA performed is the non-specific, fixed versus random t-test. We choose $N=200$ for the number of traces in each group, for a total of 400 traces per TVLA performed. This number of traces creates large separation between points of low leakage and ones of high leakage, as seen in Figure \ref{TVLA_Surface}(a), where the high leakage location reaches a t-value of 22, while the low leakage location only reaches a t-value of 4. Note that the TVLA surface is rough, \textcolor{black}{with many local minima and maxima}.  Even at a fixed location there is variance in the TVLA measurements, shown in Figure \ref{TVLA_Surface}(b). However, it is infeasible to perform many TVLA measurements at each point to average out this noise.
\subsection{SNR for Leakage Measurement}
\textcolor{black}{Compared to amplitude and TVLA, SNR, as defined in~\cite{mangard_hardware_2004} requires more traces, however has a direct relationship to the MTD. Given this relationship, one can estimate the MTD, thus a location maximizing SNR will minimize MTD. 1000 traces were used to calculate the SNR, as for the 8-bit microcontroller used, this gave large separation between locations of high and low leakage, as seen in Figure \ref{SNR_Surface}, where the SNR varies from -30dB to 3dB. SNR is calculated using the same intermediate variable and leakage model as the CEMA used, that is, the first round sbox output and the the Hamming Weight model, respectively.  Like with TVLA, the surface is somewhat rough, but again it is infeasible to take many SNR measurements to average out this noise.}

\subsection{Correlation among Amplitude, TVLA, \textcolor{black}{SNR}, MTD}
\textcolor{black}{While signal amplitude, TVLA, and SNR can all be used with \texttt{SCNIFFER} as measures for leakage, since the end goal of the \texttt{SCNIFFER} system is to perform an attack, we investigate how these measures compare to the MTD at each location.
To compare the measures, a $10 \times 10$ scan of the chip was carried out, and CEMA was performed using 1,000 traces at each point. The resulting heatmap, along with heatmaps for SNR, TVLA, and amplitude, are shown in Figure \ref{MTD_Compare}. From these results, clearly TVLA and SNR both appear to correlate to the MTD strongly, however amplitude correlates very poorly. While signal amplitude is \textcolor{black}{easy to measure}, there is no guarantee that this measure correlates to the MTD, as high signal leakage does not imply high information leakage. Additionally, an uncorrelated EM source having high signal leakage could confuse an attacker into choosing a poor location to attack. While TVLA also does not guarantee high \textcolor{black}{exploitable} leakage, it can be used to identify and \textcolor{black}{focus on regions where leakage is detected with confidence}. Additionally, for the microcontroller considered in this work, TVLA does empirically correlate to the MTD quite well, even if it is not guaranteed to be the case in general. Finally, as SNR is directly related to the attack success rate, it unsurprisingly is highly correlated in practice. Further, due to this correlation,  the location of highest SNR will theoretically be the location of lowest MTD, achieving \texttt{SCNIFFER}'s goal.}

\section{Greedy Gradient-Search Heuristic}
A critical piece of the \texttt{SCNIFFER} system is the algorithm for locating the point \textcolor{black}{of high leakage at which the attack should be performed}. It is through this algorithm that the \texttt{SCNIFFER} attack framework gains benefits over an exhaustive search, as the \textcolor{black}{high leakage} location in an $N\times N$ grid can be found with $N$ measurements as opposed to $N^2$. As an example, we use \textcolor{black}{SNR as the leakage measure} to demonstrate the performance of the \texttt{SCNIFFER} greedy gradient-search algorithm throughout this section.  The remainder of this section describes the algorithm in detail, and provides results of running the algorithm on an Atmel XMEGA 8-bit processor running software AES. 

\subsection{Algorithm Description}
To avoid taking measurements at all possible points, we propose a heuristic search algorithm for finding \textcolor{black}{a point of high leakage} in a minimum number of scans. The search algorithm works in two phases. In the first phase, the search space is divided into an \textcolor{black}{$M \times M$ grid, where $M$ is the initial grid size parameter, and the leakage is measured at the center of each grid cell.} Then in the second phase, a gradient search algorithm is started from the point of the highest leakage found in the first phase. \textcolor{black}{The gradient is computed by measuring the leakage of the four grid cells adjacent to the current cell, then treating each measurement as the magnitude of a vector whose direction is the direction from the cell where the gradient is being estimated to the cell where the measurement was made. The average of these vectors is treated as the estimate of the gradient. The next point to measure is determined by adding a vector in the direction of the gradient with a magnitude of stepSize to the current location. This location is then mapped to a grid cell, and the leakage is next measured in the center of this resulting grid cell. Given this method, movement is restricted to be between grid cells, and is not entirely arbitrary, however movement to diagonal cells or moving multiple cells at once are possible moves, depending on the stepSize parameter.}

 The algorithm will stop if it attempts to measure outside the search space, instead moving only to the edge. A maximum number of iterations can also be specified, along with an ``iterations without improvement" stopping criteria. \textcolor{black}{The ``iterations without improvement” parameter should be set to a sizeable fraction of the grid resolution N, for values too small, several iterations may pass without improvement, especially for noisy surfaces, and the algorithm may stop prematurely.} This two phase process is described in Algorithm 1.

\begin{algorithm}[!t]
\SetAlgoLined
\SetKwInOut{Input}{initialGridSize, stepSize}\SetKwInOut{Output}{Point of highs leakage}
$N=$ Grid Resolution\;
maxLeakage $=0$\;
initLocs $=$ getInitialLocations$($initialGridSize, N$)$\;
\For{loc $\in$ initLocs}{
    moveProbe$($loc$)$\;
    leakage = getLeakage$()$\;
    \If{leakage $>$ maxLeakage}{
        maxLeakage $=$ leakage\;
        startLoc $=$ loc\;
    }
}
moveProbe$($startLoc$)$\;
bestLoc $=$ startLoc\;
$m =$ startLoc\;
\While{Not Converged}{
    delta $=$ getDelta$($get4Neighbors$())$\;
    $m = m - $stepSize$*$delta\;
    moveProbe$($m$)$\;
    leakage = getLeakage$()$\;
    \If{leakage $>$ maxLeakage}{
        maxLeakage $=$ leakage\;
        bestLoc $=$ loc\;
    } 
}
\caption{Gradient Search Heuristic to find the high leakage location}
\label{Grad_search_algo}
\end{algorithm}
\begin{figure}[!t]
  \centering
  \includegraphics[width=0.48\textwidth]{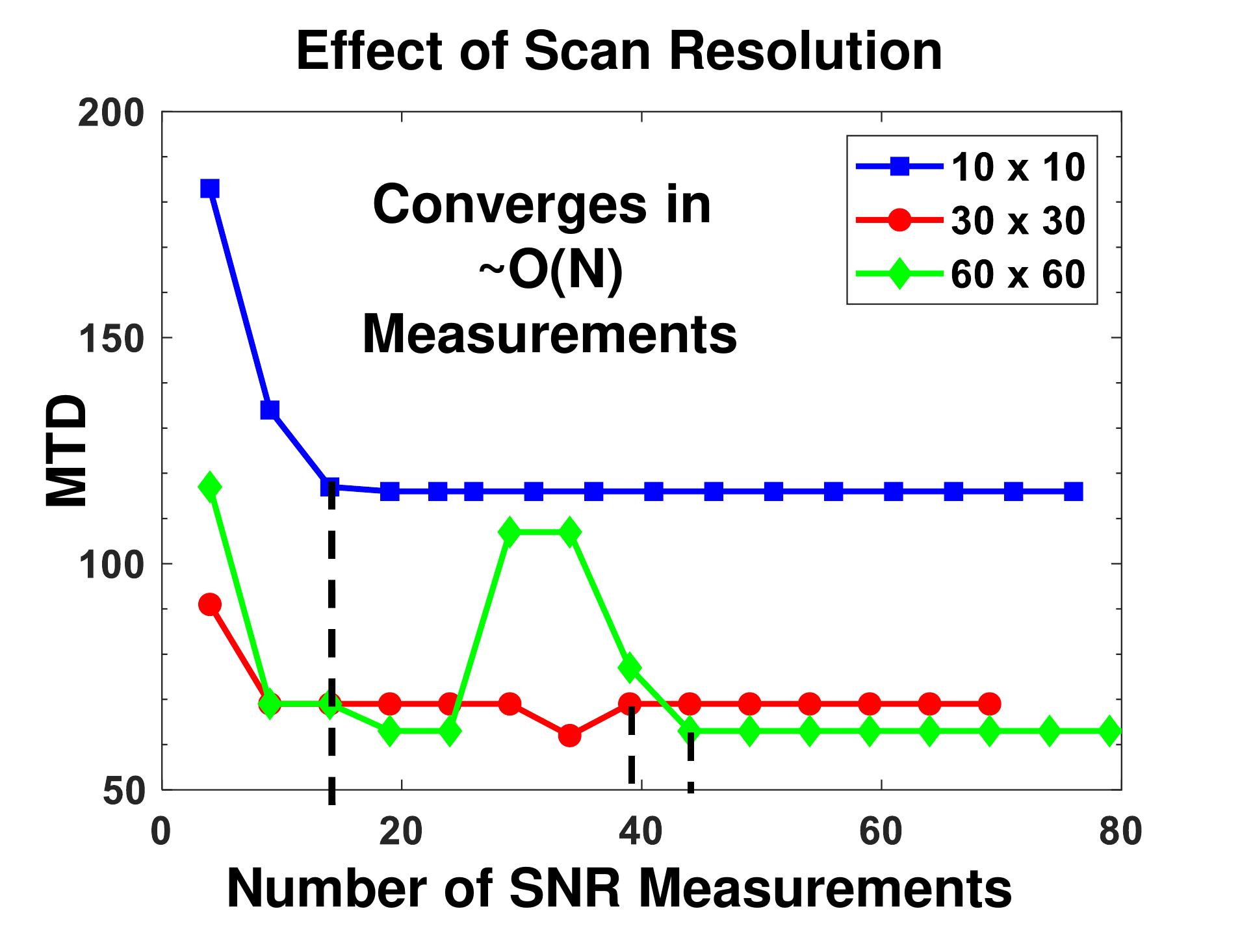}
  \caption{Leakage vs. number of \textcolor{black}{SNR} measurements for varying grid scales. The data for the $30 \times 30$ grid was the same as in Figures \ref{SNR_Heatmap} and \ref{SNR_Surface}(a). The full $60 \times 60$ and $10 \times 10$ grids were also collected, allowing the performance of the algorithm to be seen at various degrees of measurement resolution. Through these results, it can be seen that even as the size of the search space increases by $N^2$, the time to \textcolor{black}{converge} increases by only $N$.}
  \label{Grid_size_performance}
\end{figure}

\begin{figure}[!t]
  \centering
  \includegraphics[width=0.48\textwidth]{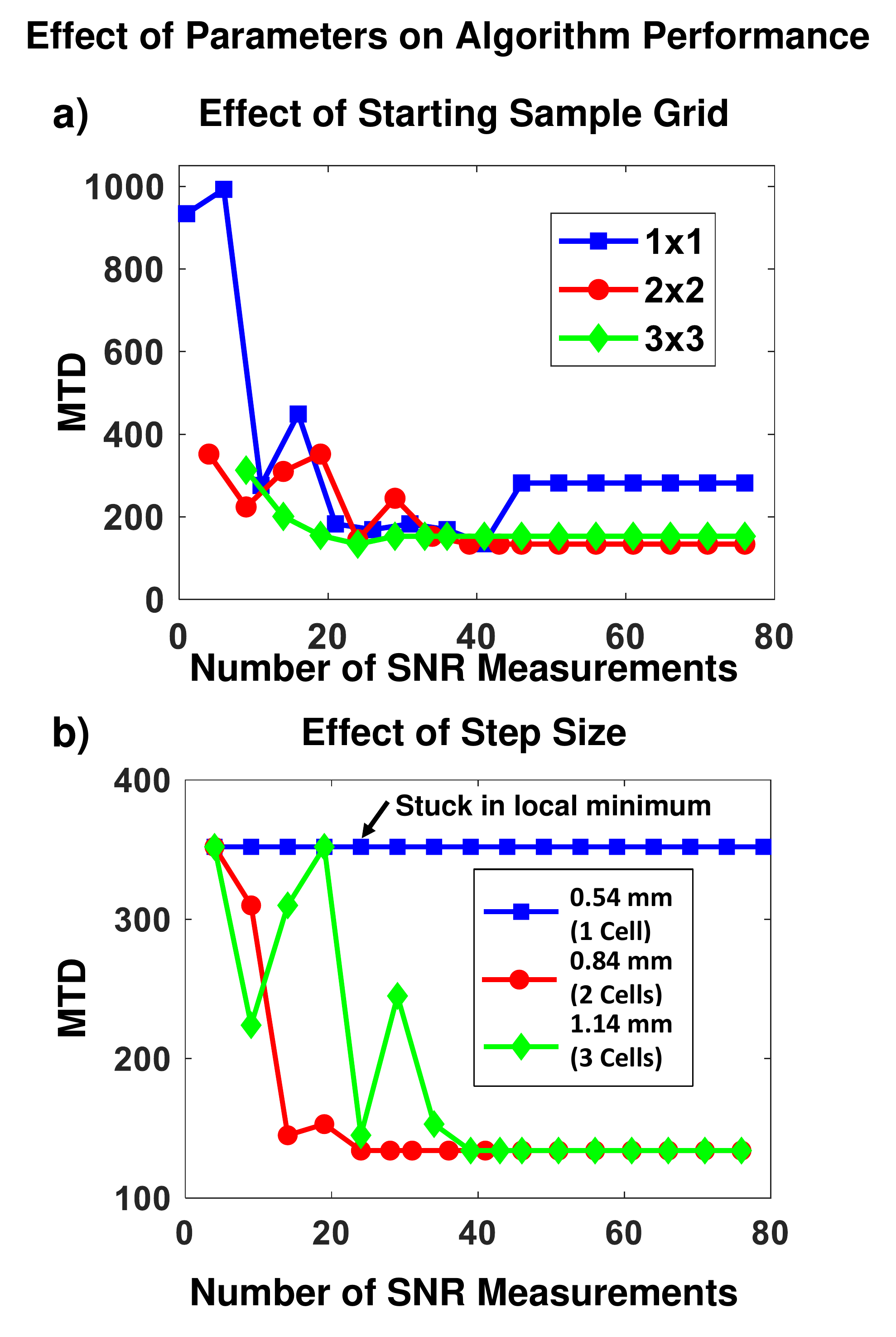}
  \caption{(a) \textcolor{black}{MTD vs number of SNR measurements performed for varying the initial sample grid size parameter. Note that the $2 \times 2$ and $3 \times 3$ grids locate the point of \textcolor{black}{high} leakage within 40 measurements, while a single point start only reaches a higher MTD, and after 45 such measurements. For all initial sample grid sizes, a step size of 1.14mm was used. (b) This demonstrates the effect of step size on performance. A step size too small can result in the algorithm getting stuck in a local maximum, and in this case as the step size increased, convergence sped up, however, for much larger step sizes, it is possible to overshoot the location of highest leakage, resulting in slower, less smooth convergence. For all step sizes, a $2 \times 2$ initial sample grid was used. Both (a) and (b) used a $30\times30$ scan resolution.} }
  \label{Parameter_performance}
\end{figure}

\begin{figure}[!t]
  \centering
  \includegraphics[width=0.48\textwidth]{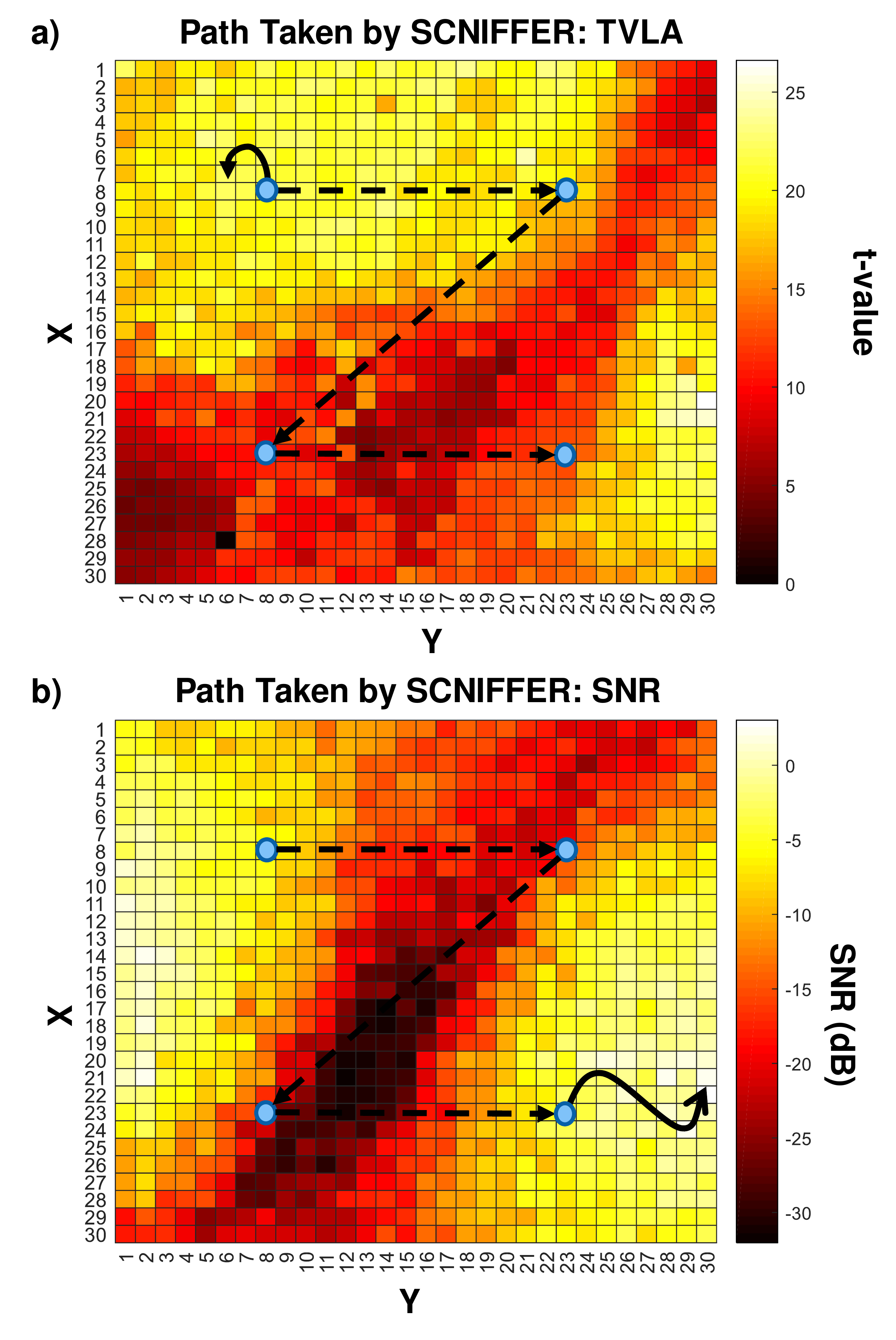}
  \caption{\textcolor{black}{Heatmaps for AES running on the 8-bit microcontroller, with the path taken by \texttt{SCNIFFER} shown for TVLA in (a), and SNR in (b). The same search algorithm parameters were used in all cases.}}
  \label{scniffer_path}
\end{figure}

\begin{figure}[!t]
  \centering
  \includegraphics[width=0.48\textwidth]{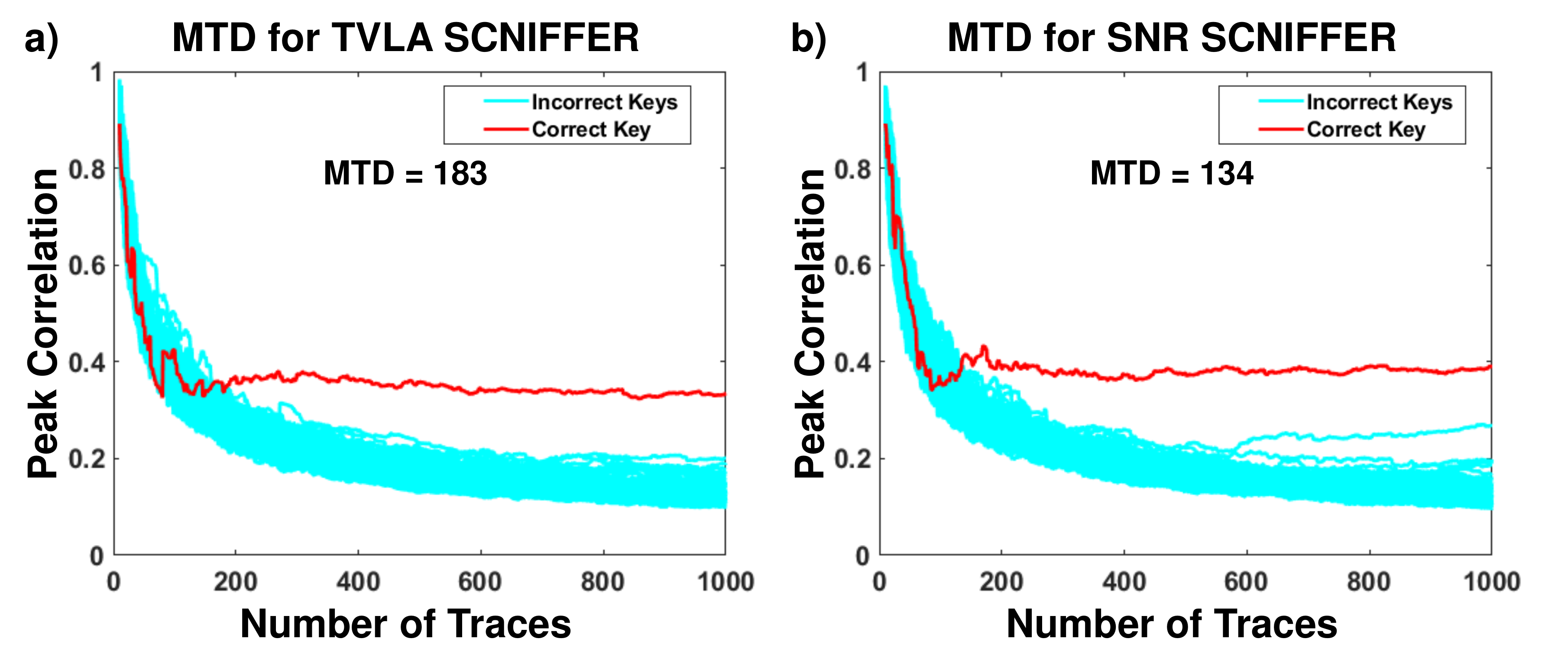}
  \caption{ MTD plots at locations found by \texttt{SCNIFFER} \textcolor{black}{using TVLA as a leakage measure (a), and SNR as a leakage measure (b). } While the MTD is not the minimum, it is \textcolor{black}{fairly close to the minimum for both measures, with SNR having a slightly lower MTD than TVLA.}}
  \label{scniffer_mtd}
\end{figure}

\subsection{Performance}
Based on experimental results, the algorithm is able to locate \textcolor{black}{a point of high leakage} in a $N \times N$ grid of possible measurements in $\approx N$ \textcolor{black}{SNR} measurements. Figure \ref{Grid_size_performance} demonstrates that as the search grid size increases by $N^2$, the number of tests required only increases by $N$, showing the improvement over an exhaustive search is more drastic as the size of the scan increases, either due to increased resolution or larger scan area. We also see the effect of the parameters of the algorithm, and see how varying them affects performance. In Figure \ref{Parameter_performance}(a), where, by increasing the resolution of the initial search grid, the lowest MTD found for a given number of measurements changes. As expected, as more initial points are scanned, fewer gradient steps are required to converge to the \textcolor{black}{high leakage location}. In Figure \ref{Parameter_performance}(b), the step size is varied, and we see that for a small step size, the algorithm \textcolor{black}{gets stuck in a local minimum, and does not converge to the point of high leakage the other step sizes do. It is worth noting that even though the algorithm gets stuck in a local minimum, the initial grid search, \texttt{SCNIFFER} still finds a relatively low MTD location.  A larger step size also converges, and if the step size is too large however, the convergence is slower, and less smooth, as it may step over the best point.} Note that the effective step size is a function of both the resolution of the scan, $N$, and the step size parameter of the algorithm. This, along with the dimensions, $L$, of the chip allow calculating the effective step size as $\frac{1}{N}*L~mm*Step Size$. Given these results, one can see that for reasonable choices of parameters, the algorithm is observed to converge to \textcolor{black}{a point of high leakage} in $O(N)$ steps for an $N \times N$ grid of measurements, providing \texttt{SCNIFFER} with a significant improvement over an exhaustive search.

\section{Results}
In this section, we provide results of using the SCNIFFER framework in various scenarios. We start with the results of an attack using \textcolor{black}{TVLA, then with SNR}. Following this, we provide a short discussion of the number of traces needed in a \texttt{SCNIFFER} attack.  We then show the performance of \textcolor{black}{the TVLA and SNR based attacks}  for a variety of cryptographic algorithms. \textcolor{black}{Next, results comparing the 8-bit architecture chip used so far to a 32-bit architecture chip are shown, again for both TVLA and SNR measures. Finally, we show results showing the effects of a Masking countermeasure, using the SNR based attack.}


\subsection{TVLA Based \texttt{SCNIFFER}}
\textcolor{black}{While it is not guaranteed to correlate with MTD, TVLA can be used with the \texttt{SCNIFFER} algorithm. The path taken for this case is shown in Figure \ref{scniffer_path}(a). This  path remains in the zone of high TVLA values, and as TVLA correlates well with MTD in our experiments, this location has a very low MTD, seen in Figure \ref{scniffer_mtd}(b), and is among the lowest on the chip. TVLA at each location requires a total of 400 traces to compute TVLA\textcolor{black}{, and additional traces would be needed for systems with lower SNR, as we describe in section IV D}. Additionally, as the TVLA surface is not smooth, convergence is slightly slowed, increasing the attack time.}

\subsection{SNR Based \texttt{SCNIFFER}}
\begin{figure}[!t]
  \centering
  \includegraphics[width=0.48\textwidth]{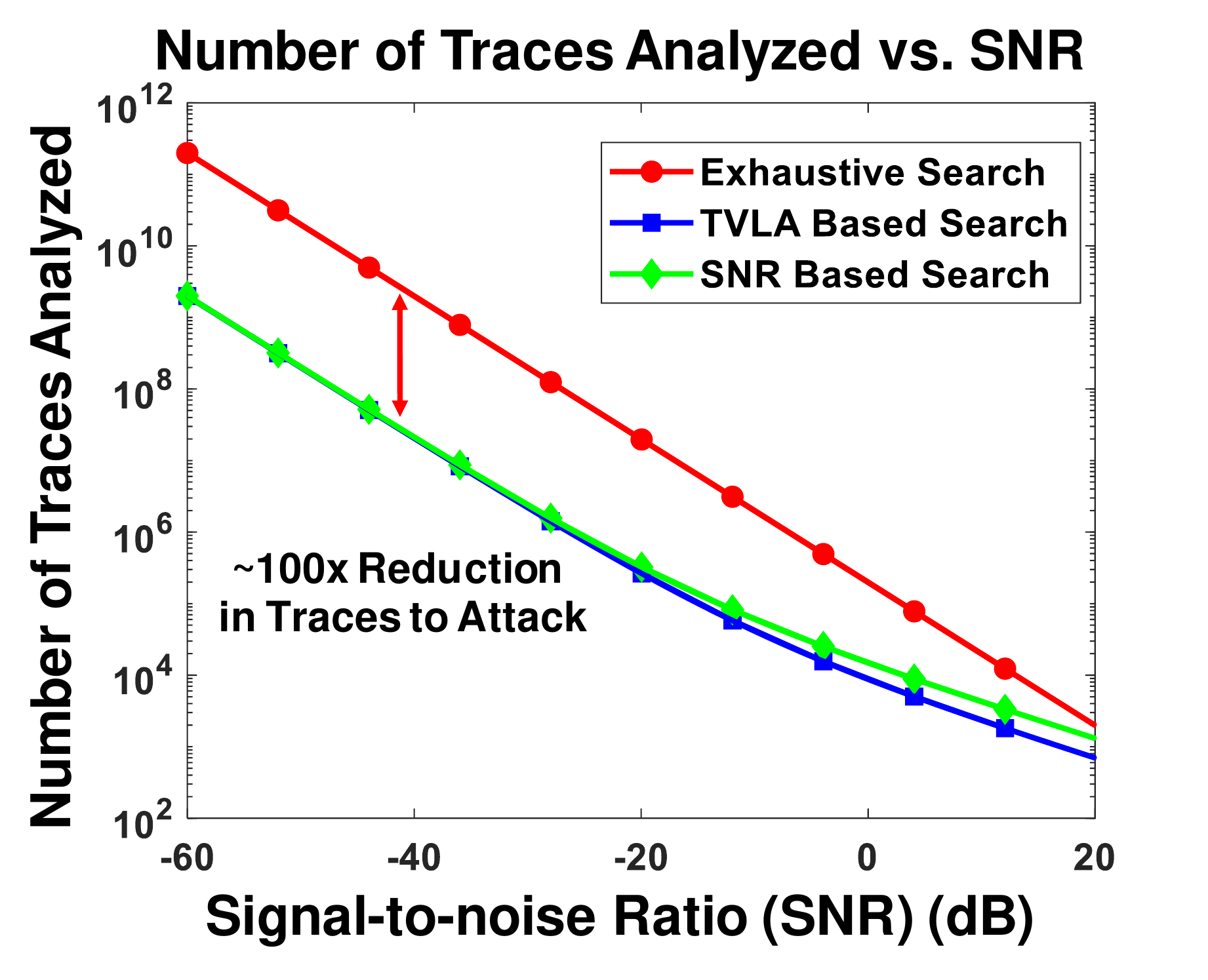}
  \caption{Number of traces required for \textcolor{black}{TVLA and SNR based} \texttt{SCNIFFER} compared to exhaustive search vs. SNR for the case of a $10 \times 10$ scan. The $\sim100\times$ reduction is due to the fact that an exhaustive search must perform a CEMA at each location, while SCNIFFER only visits $N$ locations.}
  \label{snr_analysis}
\end{figure}

\textcolor{black}{In contrast to TVLA, which does not guarantee leakage found is exploitable, SNR does, as it is related to the MTD. We see that SNR based \texttt{SCNIFFER} does take a different path than TVLA, and converges to a different location. The MTD at this location is slightly lower than the TVLA location, but still not the absolute lowest found on the chip. Furthermore, to accurately measure SNR, more traces than TVLA are needed for measurement, increasing the number of traces needed, and this number increases as the SNR reduces, as discussed in section IV D. Despite this, once the SNR reduces below a certain point, shown in Figure \ref{snr_analysis}, a SNR-based \texttt{SCNIFFER} attack becomes as efficient as a TVLA-based attack, with the additional guarantee of exploitable leakage.}

\begin{table}[]
\centering
\begin{tabular}{@{}llll@{}}
\toprule
Leakage Measure  & \begin{tabular}[c]{@{}l@{}}Convergence\\ Location\end{tabular} & MTD & Total Traces \\ \midrule
TVLA  & (2, 2) & 183 & 5,807 \\
SNR  & (7, 10) & 134 & 10,134 \\
Exhaustive  & (3, 6) & 91 & 100,000\\ \bottomrule
\end{tabular}
\caption{Comparison of different \textcolor{black}{leakage measures} used with \texttt{SCNIFFER}, \textcolor{black}{as well as results of a full exhaustive search}. The total traces includes the traces needed for the initial search, gradient search, and CEMA. \textcolor{black}{The exhaustive search total traces includes a 1000 trace CEMA at all 100 locations.}}
\label{tvla_amp_comp_table}
\end{table}

\subsection{Number of Traces Needed For \texttt{SCNIFFER} Attacks}
The performance of the \texttt{SCNIFFER} platform can be quantified and compared to other methods by investigating how the total number of traces needed to perform an attack changes as the SNR of the device under attack changes. Previous works have shown in~\cite{standaert_overview_2006} and~\cite{mangard_hardware_2004} that the MTD for a CEMA attack is related to the SNR of the signal used in the attack by $MTD=k_0*\frac{1}{SNR^2}$. Additionally,~\cite{roy_cc_2019},\textcolor{black}{~\cite{roy_leak_2016}} have shown that the number of traces needed to perform a TVLA \textcolor{black}{ ($N_{TVLA}$) or calculate SNR ($N_{SNR}$)} is also related to SNR by $N_{TVLA}=c_0*\frac{1}{SNR}$ and  $N_{SNR}=c_1*\frac{1}{SNR}$. From there, it is straightforward to quantify the performance of an exhaustive search and \texttt{SCNIFFER} using both \textcolor{black}{TVLA and SNR as follows,}

\begin{equation}
N_{SCN-TVLA}=N*c_0*\frac{1}{SNR}+k_1*\frac{1}{SNR^2}
\label{NTVLA_EQN}
\end{equation}

\begin{equation}
\textcolor{black}{
N_{SCN-SNR}=N*c_1*\frac{1}{SNR}+k_1*\frac{1}{SNR^2}} 
\label{NSNR_EQN}
\end{equation}

\begin{equation}
N_{exh}=N^2*k_1*\frac{1}{SNR^2}
\label{NSOTA_EQN}
\end{equation}
where $N \times N$ is the resolution of the grid scan\textcolor{black}{, and $k_0$, $k_1$, and $c_0$ are arbitrary constants chosen such that the models match the results presented}.

A \texttt{SCNIFFER} attack requires measurements to be made at approximately $N$ points for an $N\times N$ grid, as the search algorithm requires $O(N)$ measurements, with each \textcolor{black}{requiring $N_{TVLA}$ in the TVLA case and $N_{SNR}$ in the SNR case}. Additionally a single CEMA attack requiring $MTD$ traces is needed, resulting in equations \eqref{NTVLA_EQN} and \eqref{NSNR_EQN}. An exhaustive search on the other hand would require a CEMA to be performed at all $N^2$ locations, resulting in equation \eqref{NSOTA_EQN}. These trends are pictured in Figure \ref{snr_analysis}, which clearly shows the $100\times$ reduction in required traces in the case of a $10\times 10$ scan for low values of SNR. This reduction can be explained by the fact that the number of traces needed to measure TVLA \textcolor{black}{or SNR} changes as $\frac{1}{SNR}$, compared to the MTD which changes as $\frac{1}{SNR^2}$. Additionally, the number of points traversed is only $N$, as opposed to $N^2$ for an exhaustive search. Also, we see \textcolor{black}{TVLA slightly outperforms SNR in terms of number of traces needed to perform an attack when SNR is high. For low SNR, the performance of both measures is mostly equivalent, as the number of traces needed is dominated by the CEMA, and using SNR as the leakage measure gives guarantees on the success rate of the CEMA, which TVLA does not.} 

\begin{figure}[!t]
  \centering
  \includegraphics[width=0.48\textwidth]{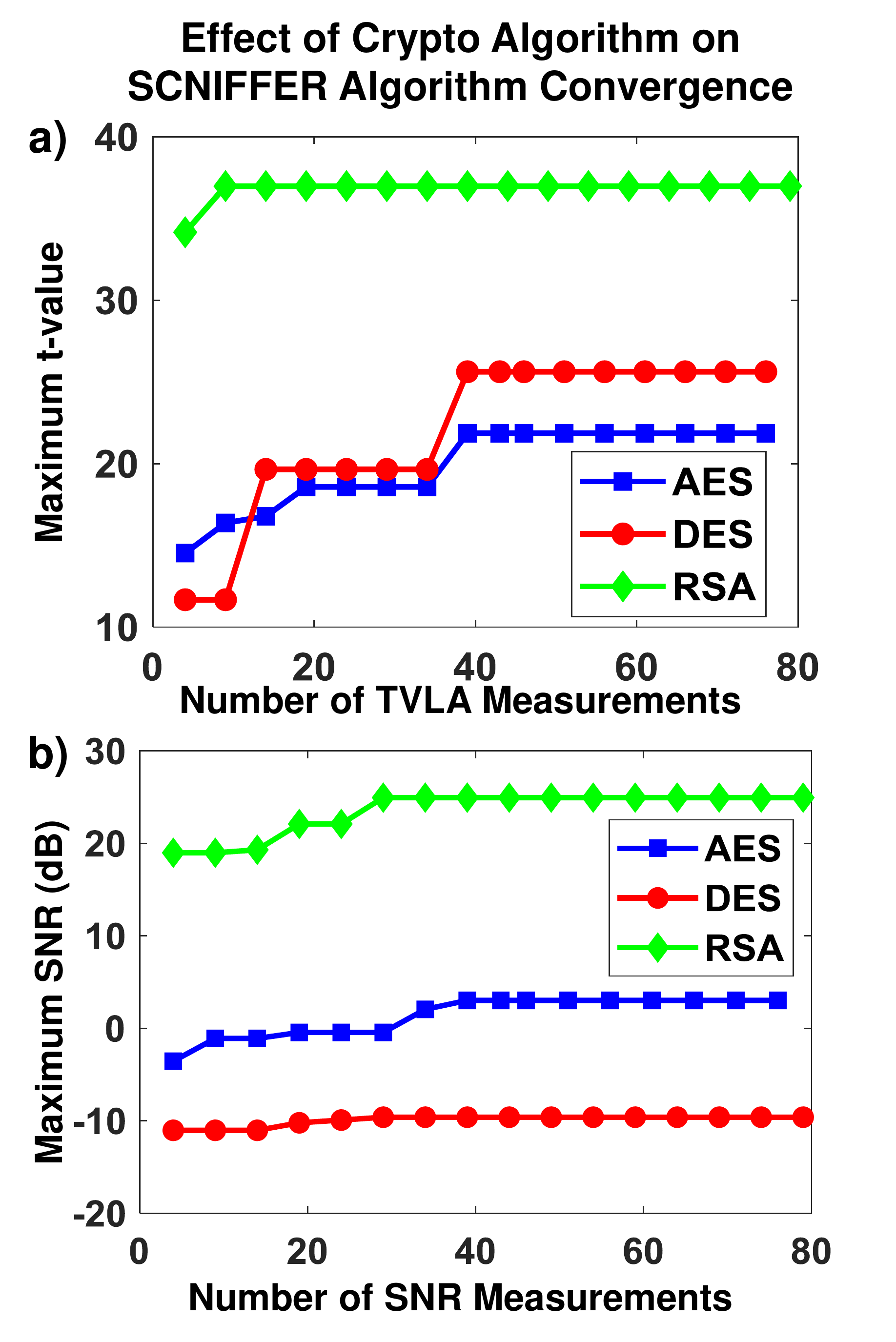}
  \caption{(a) Max t-value vs. number of TVLA tests performed for all cryptographic algorithms (AES, DES, RSA), showing the scanning algorithm performs well, finding the point of max leakage within 40 TVLA tests in all cases, \textcolor{black}{with a grid size of $30\times30$}. The initial sampling grid was $2 \times 2$ and the step size was 0.84mm. Note that for RSA, one of the initial samples is already close to the maximum, and this maximum is found in just one step. For AES and DES, whose leakage patterns are less smooth, and have smaller areas of high leakage, the time to \textcolor{black}{converge} is higher. \textcolor{black}{(b) Max SNR vs. number of SNR measurements for all algorithms (AES, DES, RSA). The search algorithm again performs well, converging in all cases in about $O(N)$ measurements. }}
  \label{algo_compare}
\end{figure}

\begin{figure}[!t]
  \centering
  \includegraphics[width=0.48\textwidth]{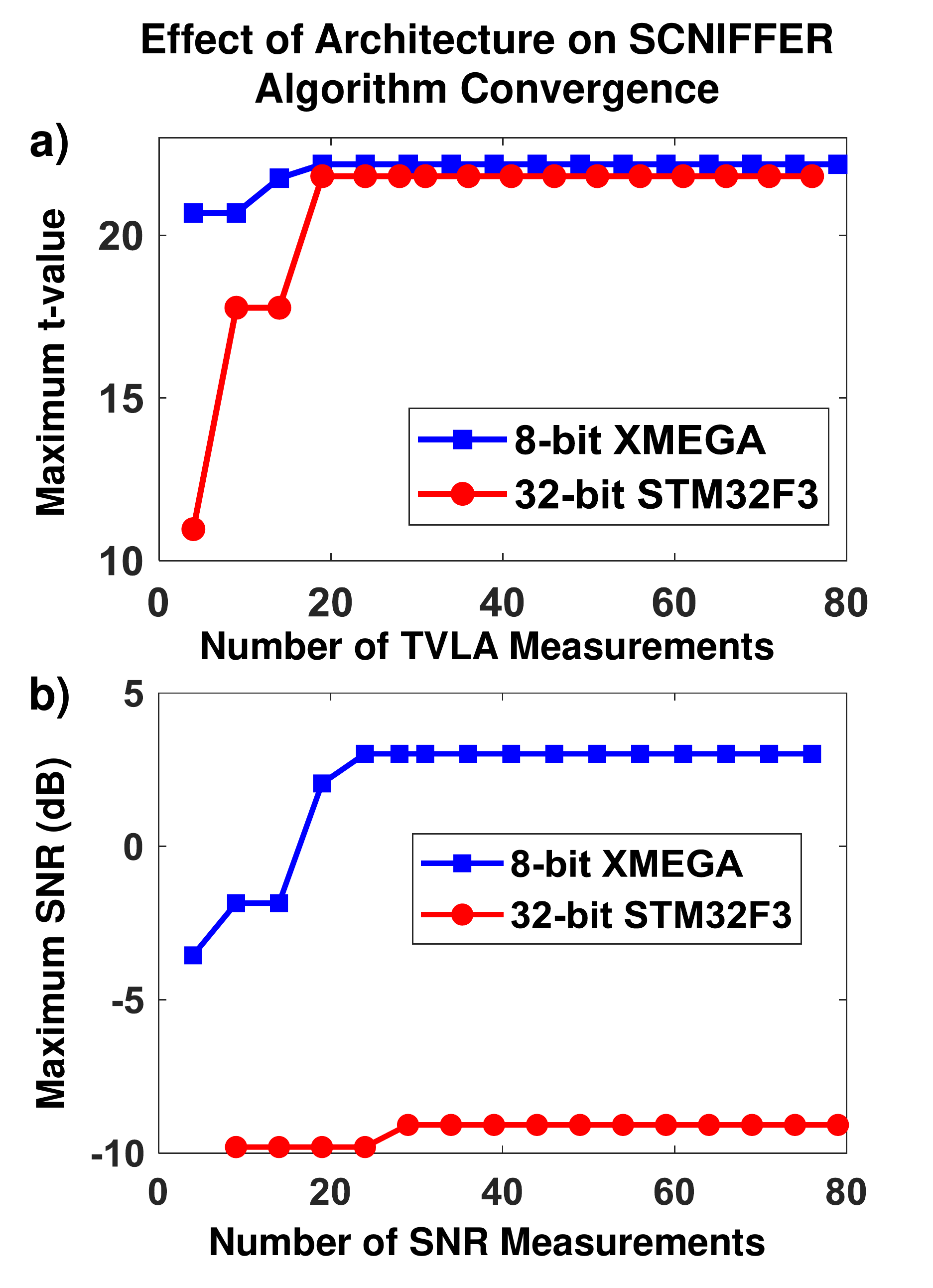}
  \caption{\textcolor{black}{(a) Max t-value vs. number of measurements for both the 8-bit XMEGA microcontroller and the 32-bit STM32F3 microcontroller. The algorithm converges within $O(N)$ measurements, where $N=30$ in both cases. the algorithm parameters used are the same as in Figure \ref{algo_compare}. (b) Max SNR vs. number of measurements for both microcontroller architectures, again showing convergence in $O(N)$ measurements. The parameters used are the same as those in part (a). }}
  \label{arch_compare}
\end{figure}

\begin{figure}[!t]
  \centering
  \includegraphics[width=0.48\textwidth]{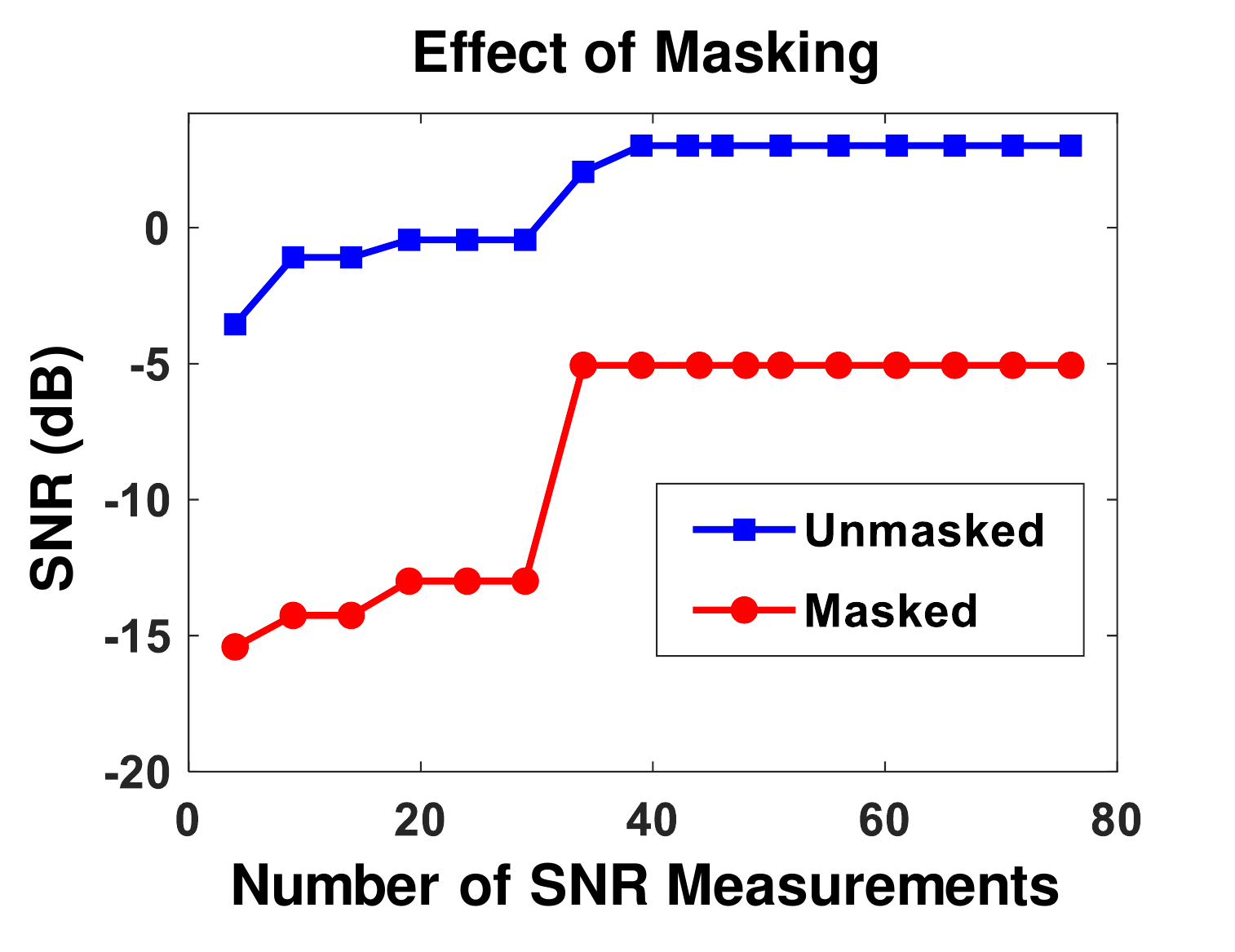}
  \caption{Max SNR vs. number of measurements for the unmasked and a masked implementation of AES on the 8-bit microcontroller. The algorithm converges within $O(N)$ measurements, where $N=30$ in both cases. the algorithm parameters used are the same as in Figure \ref{algo_compare}.}
  \label{masking_effect}
\end{figure}

\subsection{Effect of Cryptographic Algorithm on Convergence}
Next, in Figure \ref{algo_compare}(a), the effect of different cryptographic algorithms running on the target microcontroller can be seen, when using \textcolor{black}{TVLA}. For AES, DES, and RSA, the gradient search algorithm converges \textcolor{black}{a  point of high leakage} in a similar number of traces. A $30\times30$ scan was performed for all algorithms, and the parameters were fixed at a $2\times2$ starting grid and step size of 0.54 mm for all cases. A similar plot, using the same parameters but \textcolor{black}{SNR as opposed to TVLA} can be seen in Figure \ref{algo_compare}(b). Again, the search converges in approximately the same number of measurements for all algorithms. Through this, we see that the greedy gradient search algorithm performs well regardless of the specific cryptographic algorithm, and regardless of the leakage measure chosen.

\subsection{Effect of Architecture on Convergence}
Additionally, we investigate the effect of different architectures (microcontrollers) on \texttt{SCNIFFER}. Up to now, the results shown have been obtained with an 8-bit XMEGA microcontroller. We now use a 32-bit STM32F3 microcontroller running software AES as the target device. \textcolor{black}{The STM32F3 uses the same clock frequency as the 8-bit XMEGA, 7.37MHz, and sampling is again done at $4\times$ this frequency. Similarly the amplifier gain is the same as the 8-bit case. }Given the same parameters for the greedy gradient search, the \textcolor{black}{algorithm converges to a location of high leakage} within $N$ measurements, with $N=30$ in this case. These results are shown in \textcolor{black}{Figure \ref{arch_compare}(a) for TVLA, and Figure \ref{arch_compare}(b) for SNR.} In both figures, the 8-bit and 32-bit architectures are compared, given the same measurement and search algorithm parameters. In this context, it is worth mentioning that as the size of the chip under attack increases, finding the location of the cryptographic engine could be a difficult task. \textcolor{black}{In scenarios such as attacking large systems, the \texttt{SCNIFFER} framework would be extremely useful in efficiently determining the position of high leakage and then performing the attack at that point}.

\subsection{Effect of Masking on Convergence}
\textcolor{black}{Lastly, the effects of a masking countermeasure with a fixed mask on the performance of \texttt{SCNIFFER} have been investigated. We again use the same measurement and search parameters, and for both cases, the \texttt{SCNIFFER} algorithm converges in approximately $O(N)$ measurements. These results are shown in figure~\ref{masking_effect}, where we see the algorithm converges after 35-40 measurements for both masked and unmasked implementations. As one would expect, the SNR for the masked implementation is significantly lower than the unmasked implementation, but the \texttt{SCNIFFER} search algorithm is still able to locate a higher SNR region through gradient search. While the measurement parameters used here were the same as elsewhere, an important note is that for countermeasures that reduce the SNR more drastically, would require more traces to be used to calculate the SNR.}

\section{Conclusions}
This work has introduced \texttt{SCNIFFER}, a fully automated integrated system for conducting end-to-end EM side-channel attacks against cryptographic systems. \texttt{SCNIFFER} combines an EM leakage scanning platform, and correlation EM analysis into a single system, which can perform all steps of an attack automatically. The system is comprised of a low-cost custom scanning hardware and gradient search heuristic based scanning algorithm. We also plan to make our code for implementing the efficient \texttt{SCNIFFER} framework and controlling the low-cost 3-D printer for scanning publicly available.

\texttt{SCNIFFER} is capable of using a variety of measures of leakage, and the search algorithm was shown to find \textcolor{black}{a location of high leakage} in an $N\times N$ chip search space with $O(N)$ measurements, providing a significant improvement over exhaustive search, and performing all stages of the search and attack completely automatically, removing the need for expert analysis. 

 Using this fully automated attack, it is possible to efficiently find \textcolor{black}{a point of high leakage} and launch a CEMA attack at this location at the press of a button. The attack uses a minimal number of traces, for a variety of microcontroller architectures and cryptographic algorithms. Even as the size of the chip increases, or as protections \textcolor{black}{lowering the SNR, such as masking,} are put in place, \texttt{SCNIFFER} retains efficiency. Finally, we show that as the SNR of the system under attack decreases, \texttt{SCNIFFER} attacks maintain their advantage over existing methods, reducing the number of traces needed by a factor of \textcolor{black}{$N$} compared to an exhaustive search, for an $N\times N$ scan of a chip.


\bibliographystyle{IEEEtran}
\bibliography{IEEEabrv,arxiv}

\end{document}